\documentclass{article}

\usepackage{jcappub} 
\usepackage[utf8]{inputenc}
\usepackage{appendix}
\usepackage{lineno}

\usepackage{subfigure}
\usepackage{comment}
\usepackage{float}
\usepackage[sort&compress]{natbib}
\usepackage{physics}
\usepackage{xcolor}
\usepackage{amssymb}
\usepackage{ulem} 
\newcommand{\unit}[1]{\,\mathrm{#1}}

\newcommand{\jaime}{\textcolor{black}}
\newcommand{\sergio}{\textcolor{black}}

\usepackage[all]{hypcap}

\title{Characterization of Atmosphere-Skimming Cosmic-Ray Showers in High-Altitude Experiments}

\author[a]{Matías Tueros,}
\author[b]{Sergio Cabana-Freire,}
\author[b]{Jaime Álvarez-Muñiz}

\emailAdd{tueros@fisica.unlp.edu.ar}
\emailAdd{sergio.cabana.freire@usc.es}
\emailAdd{jaime.alvarez@usc.es}

\affiliation[a]{Instituto de Física La Plata, CONICET-UNLP,
   Diagonal 113 entre 63 y 64 , La Plata, Argentina}
\affiliation[b]{Instituto Galego de Física de Altas Enerxías (IGFAE), Universidade de Santiago de Compostela, 15782
Santiago de Compostela, Spain}

\date{\today}

\begin{document}

\abstract{Atmosphere-skimming showers are initiated by cosmic rays with incoming directions such that the full development of the cascade occurs inside the atmosphere without ever reaching the ground. This new class of showers has been observed in balloon-borne experiments such as ANITA, but a characterisation of their properties is lacking. The interplay between the Earth's magnetic field, the long distances over which atmosphere-skimming showers develop, and the low density of the atmosphere they traverse gives rise to several effects that are not seen in downward-going cascades, and require detailed modeling. In this article, we used the latest version of the ZHAireS-RASPASS shower simulation program to tackle this problem, and dwell on the particular phenomena that arises from the peculiar environment in which these showers develop. We focus in particular on the properties of the longitudinal profile of the shower and its fluctuations as a function of cosmic-ray energy, direction and primary mass. We have also studied the phase-space of cosmic-ray arrival directions where detection in high-altitude experiments is more likely, and have found that only in a small range of directions the showers are sufficiently developed before reaching the altitude of the detector. Our results are relevant for the design of high-altitude and in particular balloon-borne experiments, and for the interpretation of the data they collect.
}

\maketitle

\section{Introduction}

\textit{Atmosphere-Skimming} (AS) or \textit{stratospheric} showers are particle  cascades initiated in the atmosphere by primary cosmic rays (and potentially neutrinos and/or photons) whose trajectory does not intersect the surface of the Earth, as sketched in figure\,\ref{fig:RASPASSGeometry}. AS showers constitute a new class of air-shower events first identified with the ANITA balloon-borne radio antenna payload \cite{ANITA:2008mzi}. In the four independent ANITA flights above Antarctica, 7 \textit{above horizon} events have been detected \cite{ANITA:2016vrp, ANITA:2018sgj, ANITA:2020gmv}. These events have reconstructed zenith angles, as seen from the altitude of ANITA at $\simeq 36\unit{km}$ above sea level, that are compatible with showers crossing the atmosphere of the Earth with no core on the ground\footnote{ANITA I, II, III and IV detected 2, 1, 2 and 2 events respectively.} and are observed nearly at the (radio) Cherenkov angle, as evidenced by the few nanosecond duration of the recorded radio pulses. 
\jaime{Moreover, the Extreme Universe Space Observatory on a Super Pressure Balloon (EUSO-SPB2) \cite{Eser:2023lck}, has recently reported the observation of Cherenkov light from candidate air shower events that are consistent with being atmosphere-skimming showers induced by cosmic rays \cite{Eser:2023lck, Cummings:2023ypo}.}

The observed radio pulse properties of the measured events are in agreement with those expected from showers developing in air and emitting coherent radiation in the MHz - GHz frequency range through the so-called geomagnetic mechanism \cite{Kahn-Lerche:1966}. The detected pulses exhibited the expected almost linearly polarized electric field in the direction of the Lorentz force, perpendicular to the magnetic field of the Earth. Due to the lack of dedicated simulations of these events their energy could not be reconstructed. 
This type of showers have the potential to be detected in other balloon-borne detectors in the planning such as PUEO \cite{PUEO} \jaime{the successor of ANITA, and the POEMMA-Balloon with Radio \cite{Olinto:2023vmx}, as well as in satellite-borne experiments like the Terzina Cherenkov detector \cite{NUSES:2023iax} on board the NUSES small-satellite mission,} and POEMMA \cite{POEMMA}. Moreover, ground-based observatories using the fluorescence technique such as the Pierre Auger Observatory \cite{Auger_FD, Auger_upward, AugerPrime} and the Telescope Array \cite{TelescopeArray}, may also observe them, \jaime{although only in favourable and limited geometries where the shower maximum is sufficiently close to the detector, at a distance typically below $\sim 50\unit{km}$}. Their detectability could extend to ground arrays of antennas exploiting the radio technique such as BEACON \cite{Southall:2022yil}, GRAND \cite{GRAND} and the Pierre Auger Observatory \cite{PierreAuger:2023gql}, among others.

 As sketched in figure\,\ref{fig:RASPASSGeometry}, atmosphere-skimming showers start developing in a region of low air density propagating into a zone of higher density closer to the ground, similarly to the case of a regular downward-going shower with zenith angle $\theta<90^\circ$. However, an AS shower can further develop into a region of the atmosphere where the density starts decreasing again. Except for geometries grazing the earth surface, AS showers propagate in a very rarefied atmosphere compared to that near ground, resulting in particle cascades that spread along the shower axis over several hundred kilometers and that can even escape the atmosphere depending on the zenith angle $\theta$ (section \ref{sec:Showers:long}). The propagation in a very low density atmosphere also alters the competition between particle interaction and decay affecting the amount of energy transferred to the electromagnetic and muonic components of the shower (section\,\ref{sec:Showers:invisible}). \jaime{Moreover, AS can exit the atmosphere and continue its development, affected only by the geomagnetic field \cite{Cocconi_magnetic, Cocconi_magnetic_erratum} and particle decay, without particle interaction or significant energy loss.}
 
 The long distance typically traveled by AS showers in the longitudinal dimension implies that there is ample time for the Earth's magnetic field to deflect the particles, creating a significant charge separation. Several effects can rise from this separation depending on the orientation of the magnetic field with respect to the shower axis. For example, near the South Pole, where the magnetic field is almost perpendicular to ground, the shower is wider in the plane parallel to the surface, and narrower in a plane perpendicular to ground (section\,\ref{sec:Showers:lat}). On the contrary, at the equator where the magnetic field is horizontal, almost parallel to the surface of the Earth, the shower is wider in a plane perpendicular to ground as particles of opposite charge deviate towards and away from the ground. In this case, particles plunging deeper into the denser atmosphere are attenuated faster than particles deflected upwards into the more rarefied air, creating a distinct charge asymmetry (section\,\ref{sec:Showers:long}). In general, in any magnetic field configuration, the lower energy charged particles can be trapped in the magnetic field and keep gyrating experiencing reduced attenuation in the rarefied atmosphere.

\jaime{Limited progress} has been done in understanding the influence that the peculiar geometry, the propagation in a rarefied atmosphere, and the significant particle deflection in different magnetic field configurations have on the properties of the showers \cite{Fuehne:2023kap}. These properties will in turn largely determine their emission in radio wavelengths \cite{Tueros_Radio_ICRC2023}, as well as in the optical and UV \cite{Cummings:2021bhg, Cummings:2020ycz}. Some general characteristics of atmosphere-skimming showers were first pointed out in \cite{Fargion:2003kn} and studied in \cite{Krizmanic:2023hvf}. \sergio{However, none of these studies were performed using a comprehensive 4D Monte Carlo simulator capable of handling detailed particle propagation in the rarefied layers of the atmosphere, including deflections in the magnetic field.}

In this work, we present the first simulations of these type of events using the ZHAireS-RASPASS \cite{Tueros_ARENA2022} simulation program,  described in section\,\ref{sec:RASPASS}. In section\,\ref{sec:Showers}, we use these simulations to characterize the longitudinal and lateral profile of the AS showers as a function of energy, zenith angle, primary type and magnetic field configuration, as well as to compute the shower invisible energy. Many of these results are relevant for the detection of these events with fluorescence light, Cherenkov light \cite{Krizmanic:2023hvf,Cummings:2023ypo} or radio waves \cite{Tueros_Radio_ICRC2023}, as well as for their reconstruction. Finally, in section\,\ref{sec:conclusions} we conclude the paper, summarizing our findings. 

\section{Simulation of Atmosphere-Skimming showers}
\label{sec:RASPASS}

\begin{figure}[H]
    \centering
    \includegraphics[trim=2.7cm 7cm 22cm 6cm, clip, width = 0.95\textwidth]
    {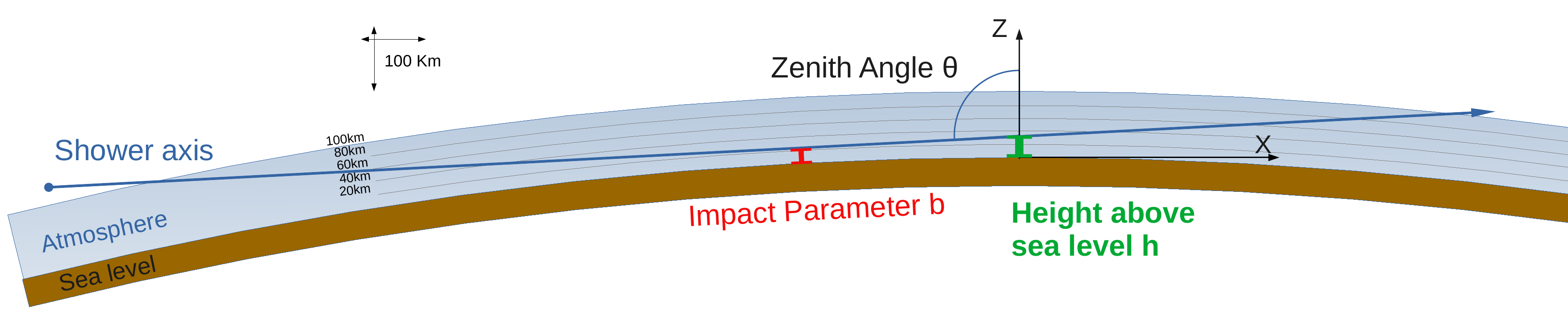}
    \caption{Sketch of the geometry of an atmosphere-skimming shower with impact parameter $b\simeq20\unit{km}$ (red line), defined as the altitude of closest approach of shower axis (represented by the blue solid line with an arrow) to ground, corresponding to a shower zenith angle $\theta=94^\circ$. The shower axis crosses the $Z$-coordinate axis at an altitude above sea level $h=36\unit{km}$ (green line). Earth's curvature is shown to scale.
}
    \label{fig:RASPASSGeometry}
\end{figure}

The geometry of Atmosphere-Skimming shower events, sketched in figure\,\ref{fig:RASPASSGeometry}, can be unambiguously characterized by the minimum (perpendicular) distance between the shower axis and the surface of the Earth, the so-called \textit{impact parameter} $b$. Alternatively, the geometry of this type of showers can be defined with respect to a high-altitude experiment located at an altitude $h$ above sea level along the vertical to the ground $\mathrm{Z}$, using the zenith angle $\theta$ w.r.t. $\mathrm{Z}$ as shown in figure\,\ref{fig:RASPASSGeometry}. Given $h$ and $\theta$, the impact parameter $b$ and hence the geometry of the shower axis is uniquely determined.

AS showers have a maximum zenith angle given by the angle at which the horizon is seen from the detector at altitude $h$. For the particular case of the ANITA balloon-borne detector at $h\simeq 36\unit{km}$ altitude, the horizon is seen at $\theta\simeq 96^\circ$, which corresponds to $b\simeq 0\unit{km}$. For $\theta\gtrsim96^\circ$ as seen from ANITA, the impact parameter is negative and the shower axis intercepts the ground without reaching the detector.

The simulation of AS showers was performed with the Monte Carlo program \textbf{R}ASPASS that stands for \textbf{A}ires \textbf{S}pecial \textbf{P}rimary for \textbf{A}tmospheric \textbf{S}kimming \textbf{S}howers \cite{Tueros_ARENA2022}. RASPASS was initially developed in 2011 as a module to simulate \textit{special} primary particles with the shower Monte Carlo simulation program AIRES \cite{AIRES}, motivated by the first above-horizon events detected by ANITA \cite{ANITA:2016vrp}. Later, it evolved into a stand-alone version of ZHAireS \footnote{ZHAireS is the AIRES program with radio emission calculation capabilities}\cite{ZHAireS}, and it now  includes several modifications to allow for the simulation of all possible shower geometries: downward-going, upward-going, and atmosphere-skimming. 

ZHAireS-RASPASS (or simply RASPASS in the following) features the same physics algorithms as the standard AIRES and ZHAireS programs, the same user-friendly input and output, and adds the capability to simulate showers initiated by multiple primaries (as for example the decay products of a tau lepton) in any event geometry. AIRES models 3D+time particle propagation in realistic conditions, incorporating atmospheric properties, geomagnetic effects, and Earth's curvature, including the possibility to study the lateral distribution of particles at any stage of shower development. The simulation includes a wide range of particles such as photons, electrons, positrons, muons, taus, pions, kaons, mesons, and baryons and nuclei up to Z = 36 (Krypton). Neutrinos are generated and their energy is taken into account, but they are not propagated. Primary particles can range from less than $1\unit{GeV}$ to over $1\unit{ZeV} = 10^{21} \unit{eV}$ in energy.

\vspace{0.15cm}
The most important physical processes considered include:
\begin{itemize}
\item Electrodynamical processes: Pair production, electron-positron annihilation, bremsstrahlung (electrons, positrons, and muons), muonic pair production, knock-on electrons, Compton and photoelectric effects, Landau-Pomeranchuk-Migdal (LPM) effect, and dielectric suppression.

\item  Hadronic processes: Inelastic collisions (hadron-nucleus and photon-nucleus), simulated sometimes using external multi-particle production models like EPOS, QGSJET, or SIBYLL, photonuclear reactions, nuclear fragmentation (elastic and inelastic).

\item Propagation of charged particles: Losses of energy in the medium (ionization), multiple Coulomb scattering, and geomagnetic deflections.
\end{itemize}

RASPASS also inherits the approximations used in AIRES/ZHAireS. In the context of this study, the most important is the use of a magnetic field with constant modulus and direction along the whole shower development. Calculations done using the International Geomagnetic Reference Field IGRF13 model \cite{IGRF13} reveal that, for the shower geometries explored in this work and an experiment located at the South Pole, the changes in the intensity of the magnetic force are less than 15\% along the development of the cascade over hundreds of km, while the change in the geomagnetic angle is less than $10^\circ$. For this reason, we expect that assuming a constant magnetic field should not significantly affect the main properties of AS showers described in this work. In locations other than the South Pole the geomagnetic field gradient is smaller, and we expect the approximation of using a constant field to be even better.

\section{Phenomenology of Atmosphere-Skimming showers}
\label{sec:Showers}

Valuable insights into atmosphere-skimming showers can be obtained by analyzing the phase-space available for shower development, before performing a detailed simulation with RASPASS, only looking the atmospheric density profile model. 

\begin{figure}[ht]
    \centering
    \includegraphics[width = .95\textwidth]{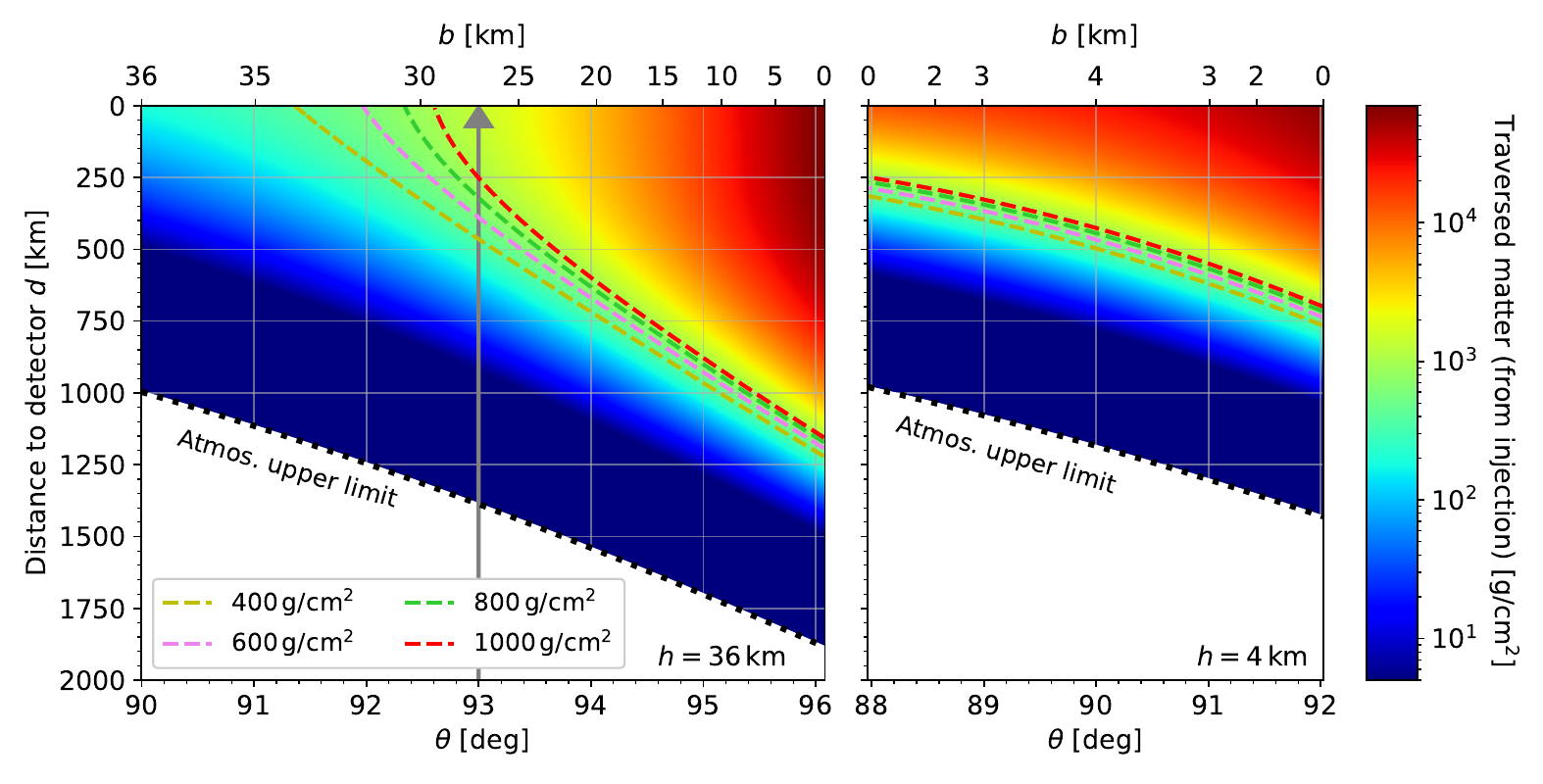}
    \caption{Phase space for the development of atmosphere-skimming air showers. The bottom (top) $x$-axis represents the zenith angle $\theta$ (impact parameter $b$) of the shower. \jaime{The left $y$-axis indicates the distance $d$ to the detector (in km), measured along the shower axis. This distance decreases as the shower evolves and gets closer to the detector, being $d=0$ by definition when the shower axis arrives at the detector. The dotted black line marks the edge of the atmosphere.} For fixed $\theta$, the colour scale represents the traversed matter, along the shower axis, from the injection point up to a point of the shower axis at a given distance from the detector. The dashed colored lines show the points where a given slant depth along the shower axis (see legend) is reached. In the left panel, the gray arrow represents the axis of a shower with $\theta=93^\circ$ developing in the longitudinal direction, entering the atmosphere at a distance to the detector $\simeq 1386\,\unit{km}$. Left panel: detector at an altitude $h=36\unit{km}$ above sea level, see figure\,\ref{fig:RASPASSGeometry}. Right panel: detector at $h=4\unit{km}$ above sea level.}
    \label{fig:PhaseSpace}
\end{figure}

In the left panel of figure\,\ref{fig:PhaseSpace} we show the integrated matter that a shower, starting at the top of the atmosphere, would cross before reaching the position of a detector in the South Pole at an altitude $h=36\unit{km}$ above sea level. The amount of atmospheric grammage (in $\unit{g/cm^2}$) given by the color scale is shown as a function of the distance to the detector along the shower axis $d$, with the detector located at $d=0$ by definition. The grammage is also shown as a function of the zenith angle $\theta$. For instance, a shower with $\theta=93^\circ$ (gray arrow in the left panel of figure\,\ref{fig:PhaseSpace}), enters the atmosphere at a distance $d\simeq 1386.5 \unit{km}$ to the detector, and has an available amount of matter to develop before the detector of $\simeq 1384.9 \unit{g/cm^2}$, enough to reach its maximum development at a depth $X_\mathrm{max}$ which is typically $< 1000\unit{g/cm^2}$ at EeV energies. 

For a zenith angle $\theta=90^\circ$ corresponding to $b=36\unit{km}$, the amount of matter available between the top of atmosphere and the detector is $\lesssim 200\unit{g\,cm^{-2}}$, not enough for full shower development. 
\jaime{This type of events provide a unique measurement of the early part of the shower development with Cherenkov light or observing the particle shower itself \cite{Krizmanic:2023hvf}. This work does not delve into an in-depth study of these partially developed showers, but shows that RASPASS could be used for this type of studies.}
For lower zenith angles (downward-going showers, not shown in the left panel of figure \,\ref{fig:PhaseSpace}),  the available mass is even lower.

Conversely, if $\theta$ increases above $90^\circ$, the shower has a smaller impact parameter $b$ and develops at lower altitudes in the atmosphere. As a consequence, the total amount of matter available for shower development increases and the shower reaches its depth of maximum development $X_{\rm max}$ farther away from the detector. This is illustrated with the dashed lines in figure\,\ref{fig:PhaseSpace} marking the positions along shower axis where the accumulated slant depth is between $600\unit{g/cm^2}$ and $800\unit{g/cm^2}$, corresponding to the typical range of $X_{\rm max}$ for showers at $\mathrm{EeV}$ energies relevant for this work. Slant depths of  $400\unit{g/cm^2}$ and $1000\unit{g/cm^2}$ are also plotted as dashed lines for reference. It can also be seen that showers with $\theta\lesssim 92^\circ$ reach their maximum development after the detector. The upper limit to the zenith angle is $\theta\simeq 96^\circ$, corresponding to an impact parameter $b \simeq 0\unit{km}$, where there is a maximal amount of matter of $\simeq 6.5\times10^4\unit{g/cm^2}$ between the top of the atmosphere and the detector and showers reach $X_{\rm max}$ at a maximal distance to the detector $d\simeq 1200\unit{km}$. \jaime{For these events, due to the large atmospheric depth, shower attenuation is severe and the chances of detecting the shower particles directly or the optical Cherenkov emission are small. However, the radio emission could still be detected, with the caveat of the effects that atmospheric refraction could have on the propagation of the pulses produced by these near-surface events, that could lead to relative time delays between wavefronts and loss of coherence depending on the frequency \cite{ANITA:2020gmv}.} As mentioned earlier showers with $\theta>96^\circ$, hit the ground and are no longer atmospheric-skimming.

The range of zenith angles where the shower would reach $X_{\rm max}$ before the detector depends on the altitude $h$ of the detector as well as on the atmospheric density model. This is illustrated in the right panel of figure\,\ref{fig:PhaseSpace} for a detector placed at an altitude of $h=4\unit{km}$ above sea level, as could be the case of ground-based experiments located on the slope of mountains such as BEACON \cite{Southall:2022yil}. Due to the lower altitude compared to the case depicted in the left panel, showers can reach $X_{\rm max}$ before arriving at the detector even for $\theta\leq90^\circ$. The horizon, as observed from the detector's altitude, corresponds in this case to $\theta\simeq 92^\circ$.

Detectors located at higher altitudes leave more of the atmosphere below them, allowing for the detection of showers with higher zenith angles above $90^\circ$ at the expense of loosing the possibility to observe showers coming from the sides almost parallel to ground or from above, as the showers run out of matter to develop before reaching the detector. Detectors at lower altitudes have their horizon closer to $90^\circ$, but can observe showers coming from above with $\theta<90^\circ$. The optimal detector altitude will depend on the experiment science objectives.

The key role played by the geometry, magnetic field and atmospheric density profile in the development of AS showers, calls for detailed Monte Carlo simulations for different zenith angles $\theta$ and magnetic field configurations. \jaime{The ANITA payload for example, has flown at different altitudes over its four separate flights, ranging from $h\simeq 35\,\mathrm{km}$ to $\simeq 40\,\mathrm{km}$. Moreover, within each flight $h$ changed with time. Detailed studies tailored to a specific experiment would require accounting for the variation in payload altitude as a function of time, and of the magnetic field as a function of payload position. In this work and for illustration, we have adopted a fixed value of $h=36\,\mathrm{km}$, and a fixed magnetic field.} 
We have used ZHAireS-RASPASS to simulate showers with $\theta = 92^\circ,\,93^\circ\,,94^\circ\,\text{and}\,95^\circ$ corresponding to impact parameters $b\simeq 32.1\,,\,27.2\,,\,20.4\,\,\text{and}\,11.6\unit{km}$ respectively. We have obtained the longitudinal development of electrons and positrons in the shower as well as the lateral development of $e^-+e^+$ and $\mu^-+\mu^+$. This is described in the following. 

The notable characteristics of the lateral and longitudinal development of AS showers have significant implications for their detection \cite{Krizmanic:2023hvf}, especially concerning the radio technique \cite{Tueros_Radio_ICRC2023}. This will be explored further in a subsequent study.

\subsection{Longitudinal development}
\label{sec:Showers:long}
\subsubsection{Density effects}

\begin{figure}[htb]
    \centering
    \includegraphics[width = .9\textwidth]{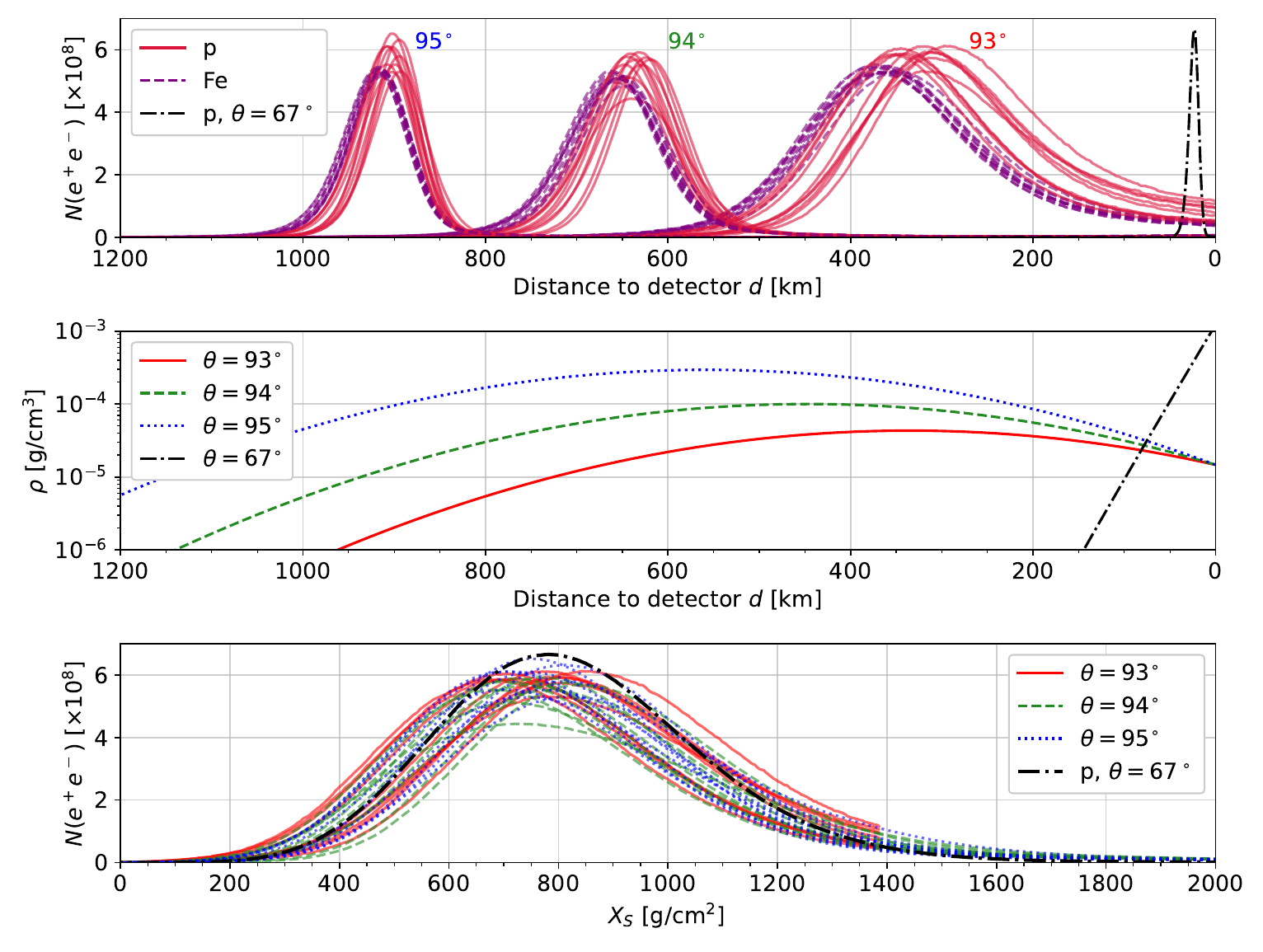}
    \caption{Top panel: Longitudinal development of the number of 
    $e^- + e^+$ of 10 proton (red, solid) and 10 iron-induced (purple, dashed) atmospheric-skimming showers of energy $E_0=10^{18}\unit{eV}$, for different zenith angles $\theta$ (corresponding to different impact parameters $b$) and $h=36\unit{km}$ - see Figs.\,\ref{fig:RASPASSGeometry} and \ref{fig:PhaseSpace}. Shower simulations were performed with ZHAireS-RASPASS with a magnetic field perpendicular to the ground plane of intensity $50\unit{\mu T}$. The longitudinal development is shown as a function of the distance to the detector along the shower axis $d$, with the detector located at $d=0$ by definition. For comparison the longitudinal development of a single proton-induced downward-going shower with $\theta=67^\circ$ is also plotted. Middle panel: Density profile of the atmosphere along the shower axis for the different zenith angles including that of the $\theta=67^\circ$ shower (dashed-dotted black line).  Bottom panel: Same as top panel as a function of the amount of matter traversed (grammage $X_s$) along the shower axis. Only the proton-induced showers in the top panel are plotted in the bottom one for clarity.}
    \label{fig:LongDev}
\end{figure}

In the top panel of figure\,\ref{fig:LongDev} we show the longitudinal profile of the number of $e^-+e^+$ in 10 proton- and 10 iron-induced AS showers for zenith angles $\theta=93^\circ,\,94^\circ$ and $95^\circ$, as a function of the distance to the detector along the shower axis $d$. In the middle panel of figure\,\ref{fig:LongDev} we show the atmospheric density profile for the same set of zenith angles as in the top panel. The density at which showers develop is typically more than an order of magnitude smaller than the atmospheric density at sea level.

As the zenith angle decreases approaching $\theta=90^\circ$, the showers develop higher in the atmosphere where the density is lower and they need to travel a larger distance to accumulate enough matter to develop. This results in increasingly longer showers stretching over hundreds of km, with the length increasing with decreasing zenith angle. This is to be compared to a downward-going proton-induced shower with $\theta=67^\circ$ also shown in figure\,\ref{fig:LongDev} whose length is of the order of few tens of km.

Simultaneously, as $\theta$ approaches $90^\circ$, the total geometrical distance between the top of the atmosphere and the detector decreases as depicted in figure\,\ref{fig:PhaseSpace}. Consequently, the showers not only stretch in the longitudinal dimension, but also develop closer to the detector as indicated by the constant-grammage dashed lines in figure\,\ref{fig:PhaseSpace}. This feature is made evident in the top panel of figure\,\ref{fig:LongDev}.

In contrast to the top panel of figure\,\ref{fig:LongDev}, in the bottom panel the longitudinal development of the number of $e^-+e^+$ in the same 10 proton-induced showers is plotted as a function of the traversed slanted depth $X_s$ in $\unit{g\,cm^{-2}}$. The longitudinal shower profiles feature a rather similar shape, independently of $\theta$, when plotted as a function of $X_s$. The profiles are also similar to that of a downward-going shower of $\theta=67^\circ$. Profiles for $\theta = 93^\circ$, stop at $X_s\simeq 1385\unit{g\,cm^{-2}}$ because the simulation stops when the shower reaches the detector. This was expected in view of the available matter for shower development at $\theta=93^\circ$ shown in figure\,\ref{fig:PhaseSpace}. Showers with $\theta<93^\circ$ are expected to reach the detector even earlier in their development for the same reason, and at $\theta\lesssim92^\circ$ the shower should reach the detector even before $X_\mathrm{max}$ is reached. This was confirmed with RASPASS showers simulations of $\theta=92^\circ$, not shown in figure\,\ref{fig:LongDev} for clarity. 

\begin{figure}[htb]
    \centering
    \includegraphics[width = .9\linewidth]{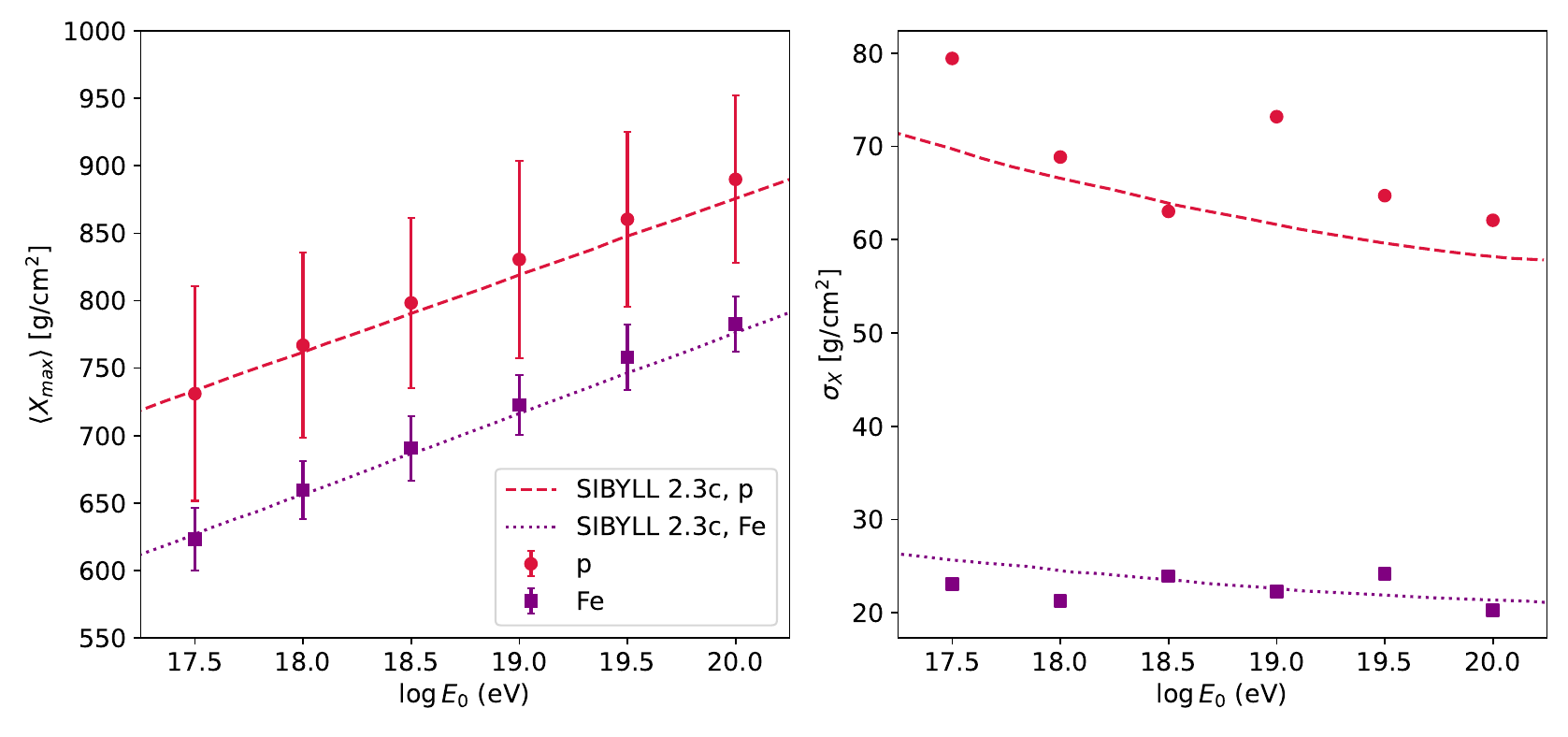}
    \caption{Left panel: Mean value of the depth of shower maximum $X_\mathrm{max}$ as a function of primary energy $E_0$, averaged over 100 proton (red) and 100 iron-induced (purple) atmosphere-skimming showers simulated with ZHAireS-RASPASS for $\theta = 94^\circ$ and a detector at an altitude $h=36\unit{km}$, using a magnetic field configuration perpendicular to the ground. The corresponding dashed red and dotted purple lines represent the $X_\mathrm{max}$ vs. $E_0$ for downward-going showers as obtained with the SIBYLL2.3c \cite{Fedynitch:2018cbl} model. Right panel: same as in the left panel but showing the \jaime{standard deviation} $\sigma_X$ of the $X_\mathrm{max}$ distribution \jaime{corresponding to the same shower simulations as in the left panel}.}
    \label{fig:Elongation_rate}
\end{figure}

Despite the fact that the low atmospheric density where AS showers develop alters the competition between interaction and decay of unstable baryons and mesons, mainly charged and neutral pions and kaons, that in turn determine many properties of the shower, the distribution of $X_{\rm max}$ and their fluctuations are roughly the same as in downward-going showers when expressed in terms of depth of matter in units of $\unit{g\,cm^{-2}}$. This is shown in the left panel of figure\,\ref{fig:Elongation_rate} where the average $X_{\rm max}$ as obtained in RASPASS simulations are plotted as a function of primary energy for AS showers, and is seen to follow closely that expected for downward-going showers.

For showers developing in very low density layers of the atmosphere, event-to-event fluctuations of a few $\mathrm{g/cm^2}$ in the depth of first interaction  and/or the depth of shower maximum, translate into differences of several tens of kilometers in the distance $d$ (relative to the detector) where the bulk of the shower is located. This effect is enhanced the lower the density of the atmosphere (the closer the zenith angle is to $90^\circ$), and leads to larger fluctuations in distance for showers with smaller $\theta$ as can be seen in the top panel of figure\,\ref{fig:LongDev}. The scale of these fluctuations depends strongly on the density profile along the shower axis, which is itself strongly dependent on $\theta$ as shown in the middle panel of figure\,\ref{fig:LongDev}. In contrast, the shower fluctuations in terms of slant depth instead of distance follow those expected for downward-going showers as shown in the right panel of figure\,\ref{fig:Elongation_rate}.

\subsubsection{Magnetic field effects}
We have also studied the key role played by the geomagnetic field on the shower longitudinal development. For this purpose, in figure\,\ref{fig:LongDev_Bfield} we show separately the longitudinal development of $e^-$ and $e^+$ for proton-induced showers and zenith angles $\theta=93^\circ$ and $95^\circ$ obtained with ZHAireS-RASPASS. To illustrate the effect of particle deflections in the Earth's magnetic field on shower development, the simulations were performed with three different magnetic field configurations namely, no magnetic field, and magnetic field of intensity $50\unit{\mu T}$ oriented either perpendicular or parallel to the ground at the position of the detector. 

In the four top panels of figure\,\ref{fig:LongDev_Bfield}, the longitudinal profiles are shown as a function of the distance along shower axis, where the shower axis is almost parallel to ground for both zenith angles shown. For the discussion that follows it is important to keep in mind that in Monte Carlo simulations (including the RASPASS program), the longitudinal profile is obtained counting the number of particle tracks crossing planes perpendicular to the shower axis placed at different fixed distances \jaime{along shower axis}. Different effects on the longitudinal shower profile are seen depending on the orientation of the magnetic field:

\begin{itemize}
\item 
 In the case of the vertical magnetic field, electrons and positrons tend to deviate in a plane that is almost parallel to the ground. The deflected $e^-$ and $e^+$ travel through similar atmospheric density profiles and their number attenuates in matter in a similar way. In this case,  the longitudinal profile of the shower is not significantly affected by particle deflections in the magnetic field as can be checked comparing to the no magnetic field case (top leftmost panel). An expected excess of electrons over positrons is seen in both the vertical and no magnetic field configurations \footnote{\jaime{The result is very similar for a magnetic field of intensity $B=50\,\mu\mathrm{T}$ parallel to shower axis, not shown in Fig.\,\ref{fig:LongDev_Bfield}.}}. This is due to the \textit{entrainment} of electrons of the medium in the shower flow due to Compton, M\"oller and Bhabha scattering as well as positron annihilation \cite{Askaryan:1962, Zas:1991jv}.

\begin{figure}[htb]
    \centering
    \includegraphics[width = .97\textwidth]{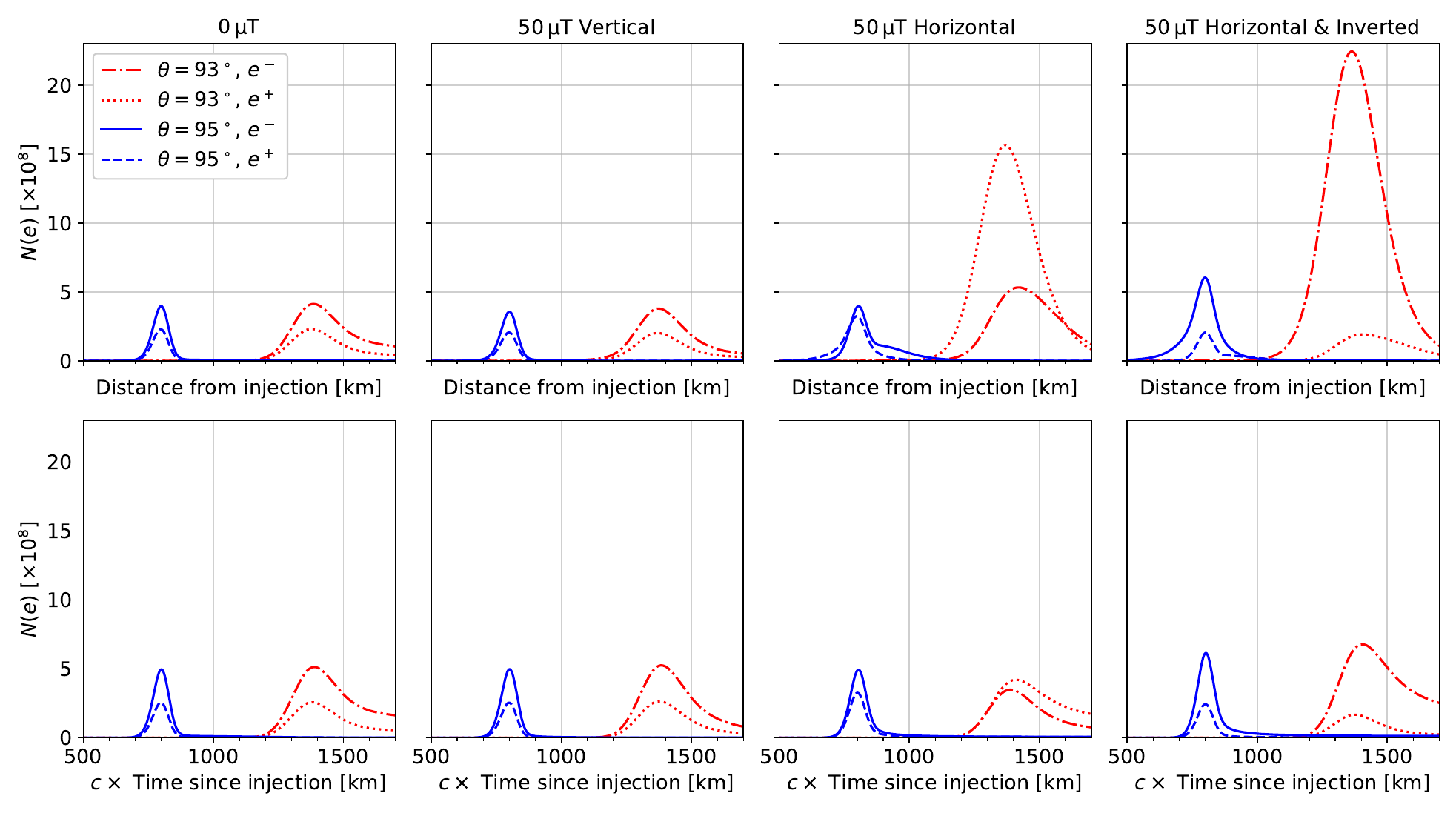}
    \caption{Average longitudinal development of the number of $e^-$ and $e^+$ over 100 proton-induced showers of energy $E_0=10^{18}\unit{eV}$, for zenith angles $\theta=93^\circ$ (red lines) and $95^\circ$ (blue lines) and the detector at an altitude $h=36\unit{km}$. Top panels: longitudinal developments plotted as a function of the distance from the entrance point of the particle in the atmosphere along shower axis; bottom panels: as a function of $ct$ with $t$ the absolute time of the shower starting at the time of injection $t=0$, representing the distance traveled by an imaginary shower front along the shower axis at the speed of light. From left to right showers were simulated with ZHAireS-RASPASS for different configurations of the magnetic field, namely, magnetic field switched off in the simulations ($0\unit{\mu T}$ - leftmost panel), and magnetic field of intensity $50\unit{\mu T}$ perpendicular to the ground at the position of the detector ($50\unit{\mu T}$ \textit{Vertical} - middle left panel), parallel to ground ($50\unit{\mu T}$ \textit{Horizontal} - middle right panel) and parallel to ground but flipped by $180^\circ$ ($50\unit{\mu T}$ \textit{Horizontal $\&$ Inverted} - rightmost panel).}
    \label{fig:LongDev_Bfield}
\end{figure}

\item 
The situation changes dramatically for the horizontal magnetic field configuration. In this case, electrons are deflected towards the ground and move to a region of increasing atmospheric density, while positrons deviate away from the ground and move to increasingly rarefied regions of the atmosphere. This asymmetry in the atmospheric density experienced by electrons and positrons significantly affects the longitudinal profile. Positrons enter regions of the atmosphere where, due to the low density of matter, they can travel large distances without scattering, annihilating or losing a large fraction of their energy through ionization. As a consequence, they have plenty of time to experience  deviations in the magnetic field, traveling in spiral trajectories depending on their energy. 

Due to these strong deflections, particles can cross planes perpendicular to shower axis several times, where they are counted in the simulation and add to the longitudinal distribution. This is the main reason for the relative increase of positrons with respect to electrons which is apparent in the top middle right panel of figure\,\ref{fig:LongDev_Bfield} when compared to the showers developing in a vertical magnetic field configuration (middle left panel), or when the magnetic field is switched off in the simulation (leftmost panel). The increase of positron counting is much more pronounced for the case of showers with $\theta=93^\circ$ than in those with $\theta=95^\circ$ because the shower develops in a more rarefied atmosphere at a greater altitude where positrons travel into even lower density regions. 

As seen in the third top  panel of figure\,\ref{fig:LongDev_Bfield} the number of counted positrons is a factor of $\simeq 3$ larger than the number of counted electrons at shower maximum. Naturally, the situation is reversed when the polarity of the magnetic field is flipped by $180^\circ$ and the shower direction is unchanged, so that electrons (instead of positrons) get predominantly deflected into the less dense regions of the atmosphere. This has been checked explicitly with a fourth set of RASPASS simulations and is shown in the top rightmost panel of figure\,\ref{fig:LongDev_Bfield}, where it is interesting to see that the electron count is even larger than in the case of positrons in the top middle right panel, because $e^-$ can travel even longer distances than $e^+$ since they do not annihilate.
\end{itemize}

\subsubsection{Longitudinal development in \textit{ct}}
The discussion in the preceding sub-section exposes the inadequacy of the traditional counting of particles in the longitudinal profile of AS showers. To address this issue, in the bottom panels of figure\,\ref{fig:LongDev_Bfield} we plot again the longitudinal distribution, but counting the number of electrons and positrons present in the shower at each time $t$. With this innovative representation of shower development, the counting of particles multiple times at a fixed plane perpendicular to the shower axis is avoided. Even for the extreme case of a particle trapped in the magnetic field, moving in a circular trajectory instead of progressing along the shower axis, the time advances and as a consequence the particle gets counted at increasingly larger values of $ct$.

As anticipated and seen in figure\,\ref{fig:LongDev_Bfield}, when plotting the shower development as a function of $ct$, the number of positrons (bottom middle right panel) and electrons (bottom rightmost panel) decreases dramatically avoiding the multiple counting and accounting for most of this effect in the simulation. However, there is still a smaller than usual negative excess charge for $\theta=95^\circ$, and even a positive excess charge for $\theta=93^\circ$ in the case of the \textit{Horizontal} magnetic field, because it is still the case that positrons lose less energy and hence live longer when they get deflected towards the lower density layers of the atmosphere. On the other hand, when comparing the top and bottom panels of figure\,\ref{fig:LongDev_Bfield} for the cases with no magnetic field $0\unit{\mu T}$, or \textit{Vertical} field, the number of particles at the maximum development shown as a function of distance $d$ (top panels) is smaller than the corresponding number of particles when the longitudinal profile is plotted as a function of time $t$ (bottom panels). The reason for this lies in the fact that in the top panels only particles crossing planes at fixed distances are counted, while in the bottom panels any particle existing at an instant of time $t$ is added in the simulation at the corresponding value of $ct$. 

Finally, it is also expected that artificially increasing the energy threshold below which charged particles are no longer tracked in the simulation, would also reduce the excess counting of positrons in the top middle-right and of electrons in the rightmost panel of figure\,\ref{fig:LongDev_Bfield}. This is because higher energy positrons and electrons suffer smaller deflections in the magnetic field and they typically have a smaller chance of crossing several times a plane at a fixed depth. We have explicitly checked with RASPASS simulations that this is indeed the case, and that the effect of excess counting decreases gradually when tracking in the simulation only particles with increasing energy thresholds. 

\subsection{Invisible energy}
\label{sec:Showers:invisible}

The very low atmospheric density in which AS showers propagate alters the competition between interaction and decay of baryons and mesons with respect to downward-going showers. In such rarefied atmosphere the interaction length of hadrons becomes comparable to its decay length at much higher energy than in conventional downward-going showers, with the transition energy increasing as the density decreases. For instance charged pions ($\pi^\pm$), that are the most numerous mesons in the shower, tend to decay over a broader energy range instead of interacting\ and this reduces the number of hadronic interactions in the shower. 

In a shower with less hadronic interactions, less secondary neutral pions ($\pi^0$) are produced and the flow of energy towards the electromagnetic component also gets reduced. In turn, the so-called \textit{invisible} energy, associated to particles that do not deposit the bulk of their energy in the atmosphere, mainly muons and neutrinos, is expected to increase with respect to that in downward-going showers. The mean energy of the muons at production is expected to increase compared to downward-going showers as $\pi^\pm$ tend to decay at higher energy. \jaime{Also, muons can eventually decay contributing to the shower electromagnetic component. This contribution is expected to be larger for those geometries in which the distance to the detector is longer, corresponding to zenith angles $\theta\gtrsim 94^\circ$ as shown in Fig.\,\ref{fig:PhaseSpace}.} These effects are expected to be enhanced the smaller the density of the atmosphere where the shower develops and, as a consequence, a strong dependence on the shower zenith angle should arise. 

\begin{figure}[ht]
    \centering
    \includegraphics[width = .9\linewidth]{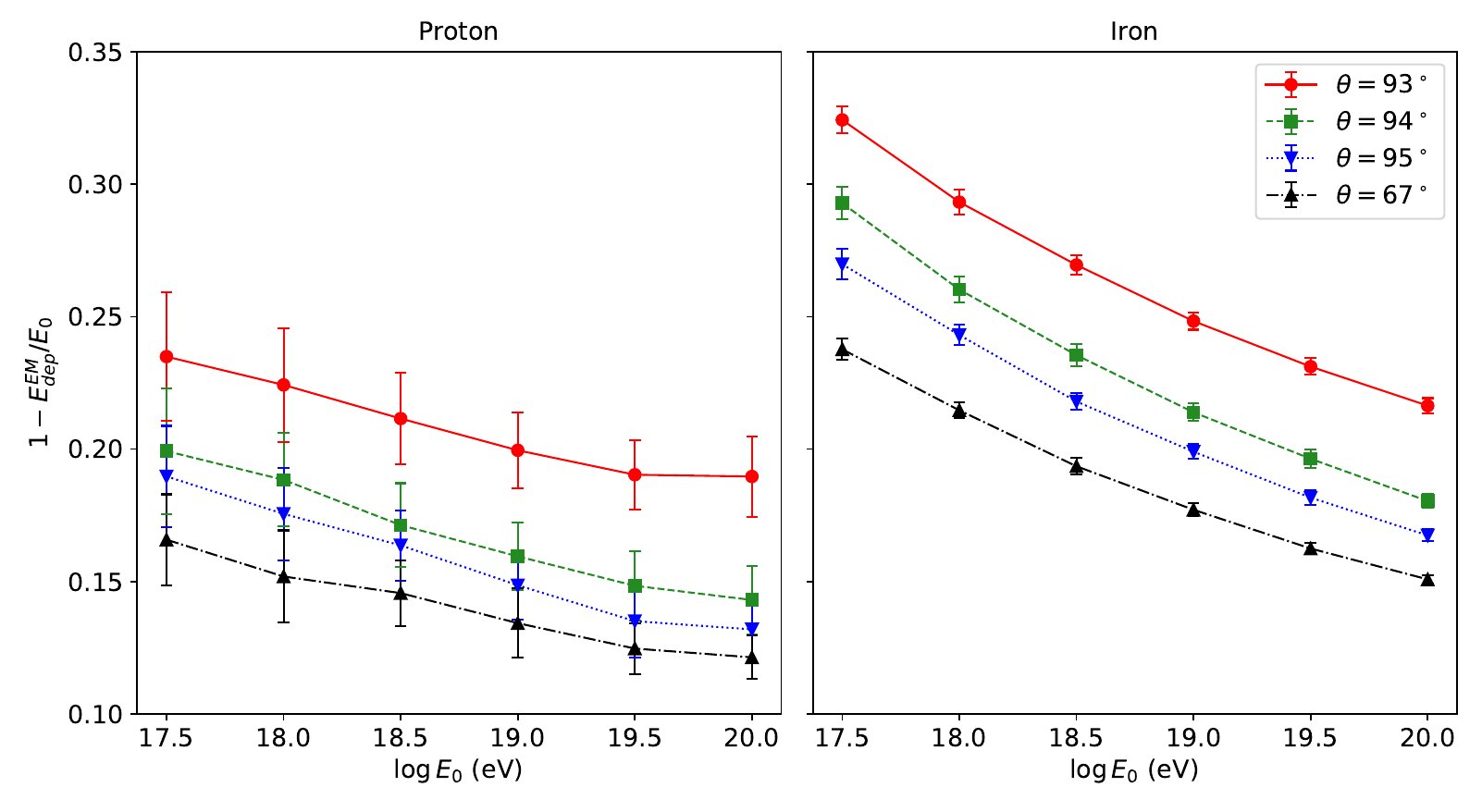}
    \caption{Fraction of \textit{invisible} energy $f_\mathrm{inv}$, defined as the energy not deposited in the atmosphere by the electromagnetic component of the shower before reaching the detector $f_\mathrm{inv}=1-(E^\mathrm{EM}_\mathrm{dep}/E_0)$, as a function of primary energy $E_0$, for atmosphere-skimming air showers induced by proton (left) and iron (right) primaries for different zenith angles $\theta$. Each marker corresponds to the average of 100 simulated showers with ZHAireS-RASPASS for a detector at an altitude $h=36\unit{km}$. For comparison, the symbols in black correspond to the average fraction of invisible energy in downward-going air showers with $\theta = 67^\circ$, also simulated with ZHAireS-RASPASS. }
    \label{fig:InvisibleEnergy}
\end{figure}

In this section we illustrate these effects, adopting the definition of invisible energy as the fraction of primary energy $E_0$ that is not deposited in the atmosphere by the electromagnetic component of the shower $f_\mathrm{inv}=1-(E^\mathrm{EM}_\mathrm{dep}/E_0)$, with $E^\mathrm{EM}_\mathrm{dep}$ the energy deposited by the electromagnetic component in the atmosphere by ionization \textit{before reaching the detector}.

In figure\,\ref{fig:InvisibleEnergy}, we show $f_\mathrm{inv}(E_0)$ in proton and iron-induced showers simulated with RASPASS for several zenith angles. $f_\mathrm{inv}$ decreases with $E_0$ as the number of generations in the shower increases and more $\pi^0$s are produced, feeding the electromagnetic cascade. The dependence of $f_\mathrm{inv}$ on the nature of the primary particle is similar to that observed in downward-going showers \cite{Auger_invisible}, with $f_\mathrm{inv}$  larger in Fe-induced than in $p$-induced showers. However, and as expected, $f_\mathrm{inv}$ in AS showers is larger than in downward-going showers, because they develop in a more rarefied atmosphere where $\pi^\pm$ decay at higher energy. This also explains the increase of $f_\mathrm{inv}$ as $\theta$ is closer to $90^\circ$ and the shower develops in a less dense atmosphere. A special case, not shown in figure\,\ref{fig:InvisibleEnergy}, is that of showers with $\theta\leq92^\circ$, because at those zenith angles and for a detector at $h=36\unit{km}$ the shower is not fully developed when it reaches the detector (see figure\,\ref{fig:PhaseSpace}). This would further reduce the amount of electromagnetic energy available to the detector, significantly increasing the experimental $f_\mathrm{inv}$. This effect is already appreciable in figure \ref{fig:InvisibleEnergy} at the highest energy for protons of $\theta = 93^\circ$, that already are developing too close to the detector. 

The behaviour of $f_\mathrm{inv}$ with $\theta$ also depends on the altitude $h$ of the detector. For the simulations performed at $h=4\unit{km}$, we have found that $f_\mathrm{inv}$ is still larger than that in downward-going showers, but smaller and exhibiting a milder dependence on zenith angle than in showers arriving at a detector at $h=36\unit{km}$. The reason for this is that these showers develop closer to the ground in a denser atmosphere where the decay of $\pi^\pm$ occurs at typically lower energies.

\subsection{Lateral development}
\label{sec:Showers:lat}

The fact that AS showers develop in the lower density layers of the atmosphere and propagate over long distances (figure\,\ref{fig:LongDev}) enhances the shower development also in the transverse direction perpendicular to the shower axis, mainly due to the fact that charged particles can travel along arcs where the gyration radius becomes comparable to the total shower length. 

\begin{figure}[htb]
    \centering
    \includegraphics[width = .9\linewidth]{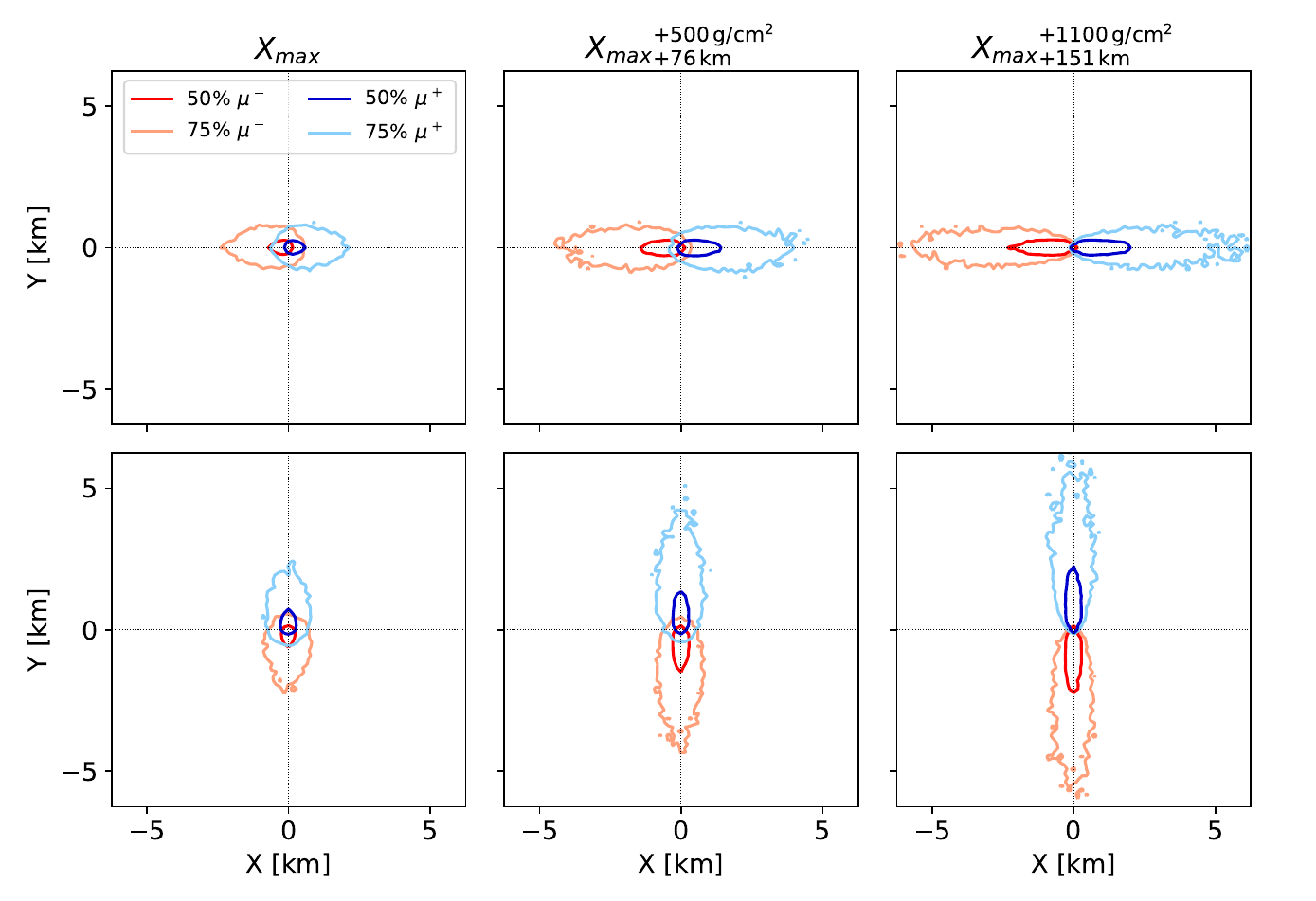}
    \caption{Lateral development of the number of $\mu^-$ and $\mu^+$ (see legend) in atmosphere-skimming showers simulated with ZHAireS-RASPASS and passing at an altitude $h=36\unit{km}$ a.s.l., with $\theta = 94^\circ$. The lateral development is shown at different stages in the longitudinal shower development, from left to right: at $X_\mathrm{max}$ \sergio{(corresponding to a slanted depth of $\sim 895\unit{g/cm^2}$ for this particular shower, reached $\sim845\unit{km}$ after injection)}, at $X_\mathrm{max}+500\,\mathrm{g\,cm^{-2}}$ and at $X_\mathrm{max}+1100\,\mathrm{g\,cm^{-2}}$. The lateral development is plotted in the directions $X$ and $Y$ perpendicular to the shower axis parallel to $Z$. The contours containing the $50\%$ and $75\%$ of $\mu^-$ or $\mu^+$ are plotted in each panel. Two magnetic field configurations were used in the simulations: perpendicular to ground along the $Y$ direction (top panels) and parallel to ground along the $X$ direction (bottom panels). }
    \label{fig:Lateral_muons}
\end{figure}

The impact of particle deflection on the lateral spread is particularly noticeable when considering the muonic component \jaime{(see further below for the case of the electromagnetic component)}, because muons travel almost without interacting, only losing energy by ionization \jaime{and eventually decaying}. This is shown in figure\,\ref{fig:Lateral_muons} where the lateral development of the number of $\mu^-$ and $\mu^+$ in showers simulated with RASPASS was obtained for two configurations of the magnetic field namely, parallel (horizontal) and perpendicular (vertical) to ground. The lateral distribution is shown, in the plane perpendicular to the shower axis, at different stages in the development of the shower. It can be readily seen that the shower is \textit{flattened} \cite{Fargion:2003kn} in the direction perpendicular to the magnetic field, with the $\mu^-$ and $\mu^+$ deviating in opposite directions and concentrating in almost symmetric \textit{lobes} of positive and negative charge. 

The lateral dimension of the shower is much larger along the direction orthogonal to the magnetic field compared to the parallel to it, and this effect is enhanced as the shower develops in the atmosphere. For instance for the particular geometry shown in figure\,\ref{fig:Lateral_muons}, the ratio of the transverse shower spread along those two directions, measuring the flattening of the shower, is a factor of $\simeq 3$ at \jaime{$X_\mathrm{max}=895\,\mathrm{g\,cm^{-2}}$ and a factor $\simeq 5$} at $X_\mathrm{max}+500\unit{g\,cm^{-2}}$.

\begin{figure}[htb]
    \centering
    \includegraphics[width = .9\linewidth]
    {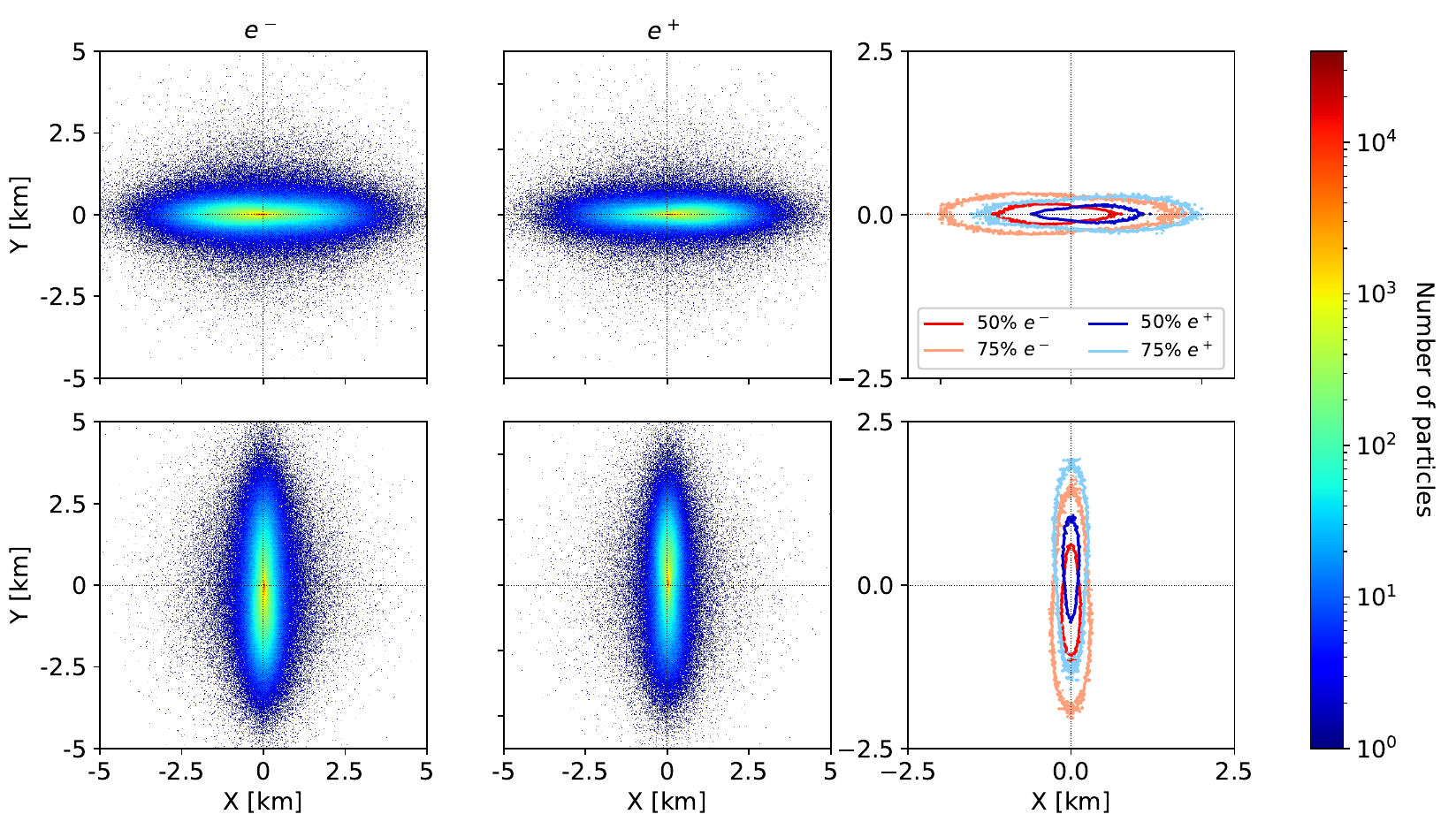}
    \caption{Lateral development of the number of $e^-$ (left panels) and $e^+$ (middle panels) in atmosphere-skimming showers simulated with ZHAireS-RASPASS and passing at an altitude $h=36\unit{km}$ a.s.l. with $\theta = 94^\circ$. The lateral development is shown at \jaime{$X_\mathrm{max}\simeq 895\,\mathrm{g\,cm^{-2}}$}, \sergio{located at a height $\sim 22.6\unit{km}$ above sea level for this particular shower,} \jaime{where the Moli\`ere radius is $\simeq 1.2\unit{km}$}. The colors indicate the number of particles following the color scale on the right. In the right top and bottom panels, the contours containing the $50\%$ and $75\%$ of $e^-$ or $e^+$ are plotted, zooming on the left and middle panels.  Two magnetic field configurations were used in the simulations: perpendicular to ground along the $Y$ direction (top panels) and parallel to ground along the $X$ direction (bottom panels).}
    \label{fig:Lateral_electrons}
\end{figure}

A similar flattening of the shower is seen in the lateral spread of the electromagnetic component shown in figure\,\ref{fig:Lateral_electrons} for the horizontal and vertical configurations of the magnetic field. As expected, there is a clear trend for $e^-$ and $e^+$ to concentrate in the regions where they are deflected by the Lorentz force namely, in the negative (positive) $X$ axis in the case of $e^-$ ($e^+$) and the magnetic field perpendicular to ground, and in the negative (positive) $Y$ axes in the case of $e^-$ ($e^+$) and the horizontal magnetic field configurations. 

The centroids of negative and positive charge are not as clearly separated as in the case of the muonic component, mainly due to multiple scattering of $e^-$ and $e^+$ and the sub-showers they generate while propagating off-axis. 
There is also a clear presence of $e^-$ (in the left panels) and $e^+$ (in the middle panels) in the directions opposite to those where they are expected to deviate in the magnetic field, for instance, in the positive $X$ axis in the case of $e^-$ and the vertical magnetic field configuration. An explanation of this effect is given in the following. 

\begin{figure}[htb]
    \centering
    \includegraphics[width = .9\linewidth]
    {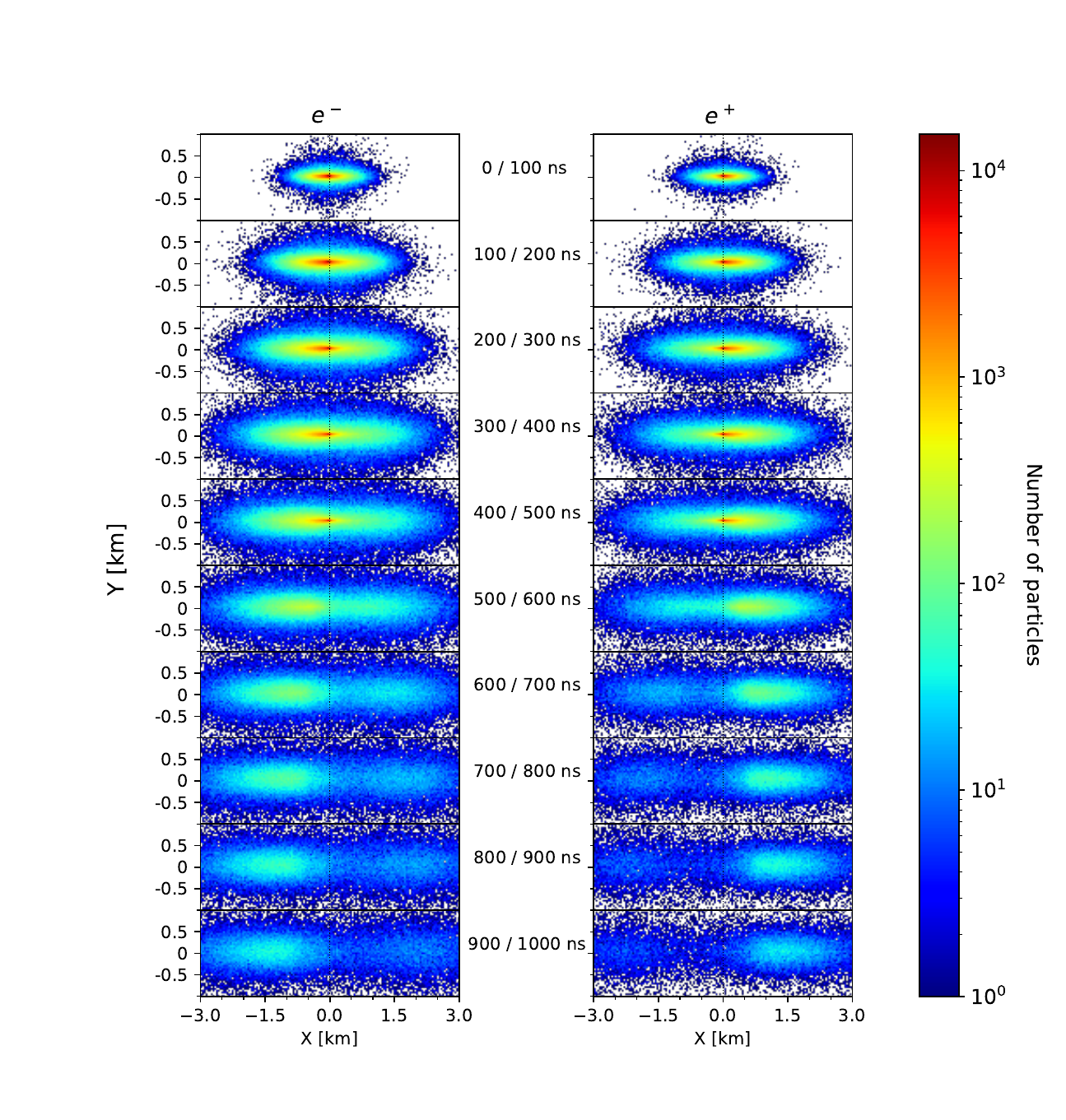}
    \caption{Time evolution of the lateral development of the number of $e^-$ (left panels) and $e^+$ (right panels) in an atmosphere-skimming shower simulated with ZHAireS-RASPASS and passing at an altitude $h=36\unit{km}$ a.s.l. with  $\theta = 94^\circ$. The colors indicate the number of particles following the color scale on the right. In all panels, the lateral development is shown at a fixed depth corresponding to \jaime{shower maximum $X_\mathrm{max}\simeq 895\,\mathrm{g\,cm^{-2}}$} but at the different time intervals shown between plots. Time $t=0$ corresponds to the arrival time at $X_\mathrm{max}$ of an imaginary particle traveling at speed $c$ along shower axis. A magnetic field configuration perpendicular to ground along the $Y$ direction was used in the simulations.}
    \label{fig:Lateral_electrons_time}
\end{figure}

In figure\,\ref{fig:Lateral_electrons} the lateral position of particles crossing, at any time $t$, a plane perpendicular to shower axis located at the depth of shower maximum $X_\mathrm{max}$ is shown. More insight into the lateral shower development can be gained by showing the particles crossing that plane at different times. This is done in figure\,\ref{fig:Lateral_electrons_time} separately for $e^-$ and $e^+$. The origin of time $t=0$ is given by the time an imaginary particle traveling along shower axis at the speed of light reaches $X_\mathrm{max}$. This novel \textit{dynamic} view of the lateral distribution of a shower allows to study the influence of the deviation of low energy electrons and positrons in the magnetic field. 

Several effects are to be noted in figure \ref{fig:Lateral_electrons_time}. As the particles spread more in the transverse dimension with $e^-$ and $e^+$ moving in opposite directions, they are also increasingly delayed in time, as expected. Furthermore, as time progresses, the centroids of charge become more distinct and prominent. 

\begin{figure}[htb]
    \centering
    \includegraphics[width = .75\textwidth]{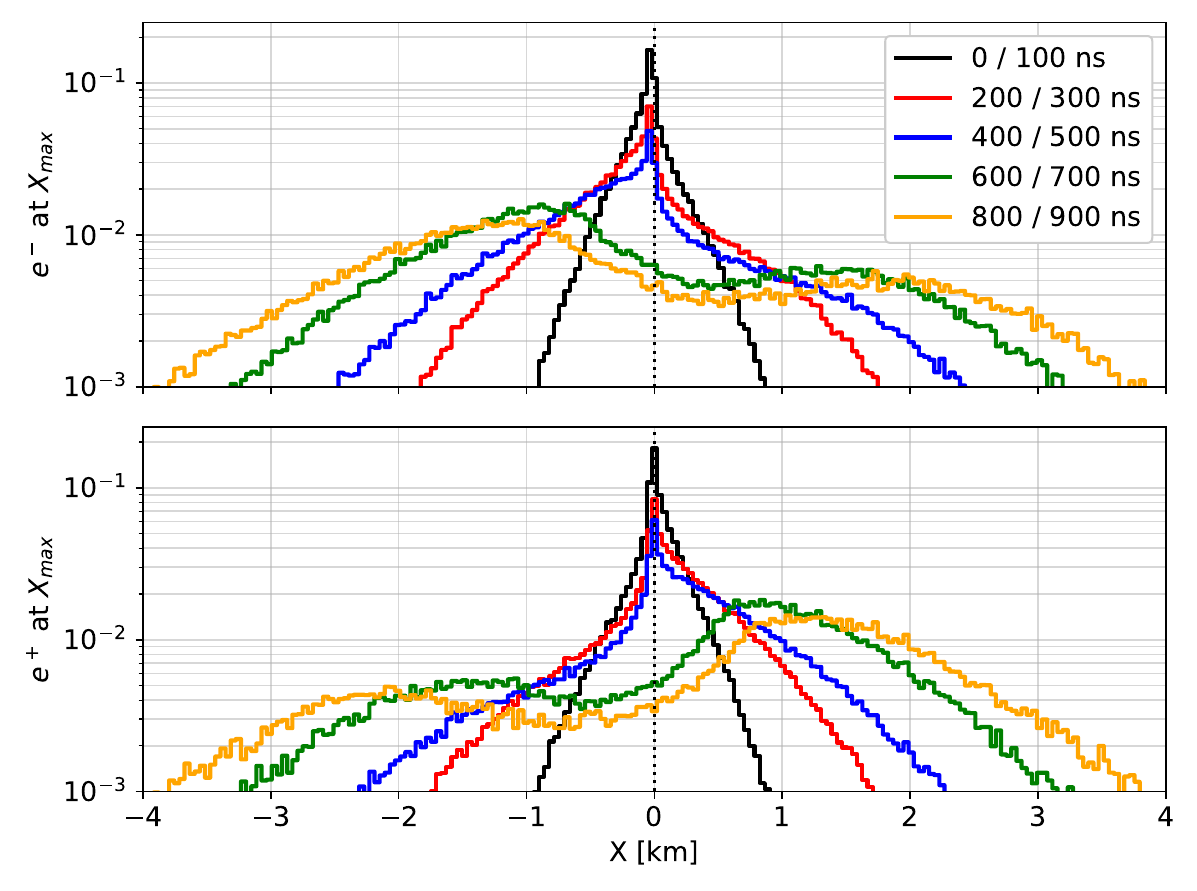}
    \caption{Projection of the lateral distribution of the number of $e^-$ (upper panel) and $e^+$ (lower panel) at a fixed depth corresponding to $X_\mathrm{max}$ and different time intervals, in an atmosphere-skimming shower simulated with ZHAireS-RASPASS and passing at an altitude $h=36\unit{km}$ a.s.l. with $\theta = 94^\circ$. The distributions shown here correspond to the projection of the histograms shown in figure\,\ref{fig:Lateral_electrons_time} along the direction parallel to ground ($X$).}
    \label{fig:Lateral_Projection}
\end{figure}

A gradual emergence of a \textit{secondary lobe} of negative or positive charge is visible, also seen previously in figure\,\ref{fig:Lateral_electrons}. As the lower energy $e^-$ ($e^+$) move towards the negative (positive) horizontal $X$-axis due to the magnetic field, they get increasing delays and cascade inducing $e^+$ ($e^-$) that cross again the plane at $X_\mathrm{max}$, contributing to the corresponding secondary lobe. In other words, cascading of positrons deflected along the positive $X$ axis in the right panel of figure\,\ref{fig:Lateral_electrons_time} generates secondary electrons that naturally appear also in the positive $X$ axis in the left panel of figure\,\ref{fig:Lateral_electrons_time}, and vice-versa. This is more apparent in figure\,\ref{fig:Lateral_Projection} where the projection along the direction parallel to ground ($X$-axis) of the two-dimensional lateral distribution plotted in figure\,\ref{fig:Lateral_electrons_time} is shown. Simulations with ZHAireS-RASPASS also indicate that this effect is less prominent for showers developing higher in the atmosphere (smaller $\theta$) because electrons and positrons cascade less in the lower density layers of the atmosphere.

\sergio{The long distances over which AS air showers develop could also allow for a direct detection of the cascade aboard high-altitude detectors \cite{Krizmanic:2023hvf}. Depending on the incoming geometry of the primary cosmic ray and the position of the detector, these measurements would sample the particle content of the air shower at very different stages of its development, allowing for unique studies of the cascade evolution. Though a detailed study of this possibility is out of the scope of this paper, the potential of this technique can be addressed using simulations with RASPASS. As an example, we show in figure \ref{fig:Lateral_Projection} the lateral distribution of $e^\pm$ produced by showers of different zenith angle passing through the position of a detector flying at an altitude $h=36\unit{km}$ a.s.l. Due to the different density profiles traversed in each case, the showers are intercepted at ages $s<1$, $s\simeq1$ and $s\gg1$ respectively (see figure \ref{fig:PhaseSpace}). The figure shows results in normal simulation conditions, following all particles until their kinetic energies are way below their rest mass; and in an artificial case where all muons with an energy below $1\unit{TeV}$ are removed, to gain insight on the contribution of muon decay to the total number of electrons and positrons at different stages of development.}

\sergio{
In the case of the shower with $\theta = 90^\circ$, which is intercepted at a very early stage of the development, the lateral distribution of $e^\pm$ barely changes when removing the lowest energy muons from the simulation. This example shows that, as expected, at ages $s\leq 1$ the electromagnetic component of the shower is mainly produced by the hadronic component through the decay of $\pi^0$'s into photons. This behaviour is also seen in the case of $\theta = 92.5^\circ$ where the shower is intercepted at $s\simeq 1$. In this case, the lateral distribution of $e^\pm$ is also wider compared to the case $\theta=90^\circ$ because the particles have been propagating under the effect of the magnetic field for a longer time.}

\sergio{
In the case of $\theta = 95^\circ$ where the shower is intercepted at a very late stage of development ($s\gg1$), several effects can be seen. The electromagnetic component is greatly reduced due to attenuation in the atmosphere and the total number of particles is much smaller than in the previous cases where $s\leq 1$. The lateral distribution is different depending on the presence or not of low energy muons, which in this geometry propagate along distances long enough to produce new electrons after losing energy through ionization and ultimately decaying. This can be seen in the right panel of figure \ref{fig:Lateral_Projection}, where lower energy muons ($E_\mu<1\unit{TeV}$), that deviate more in the geomagnetic field, are feeding the electromagnetic component away from the shower axis.}

\begin{figure}[htb]
    \centering
    \includegraphics[width = .95\textwidth]{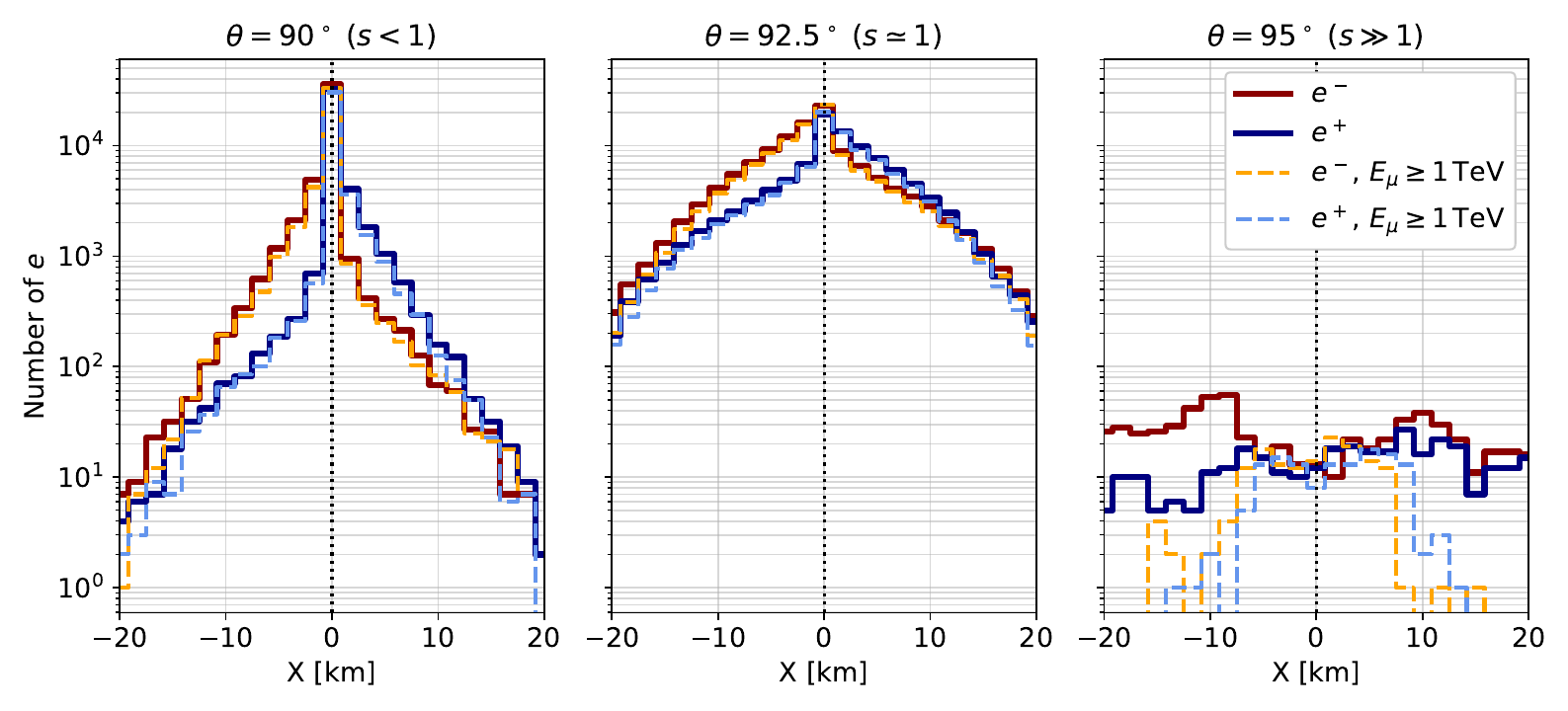}
    \caption{\jaime{Projection of the lateral distribution of the number of $e^-$ and $e^+$ in AS proton showers with $E_0=10^{18}\unit{eV}$, simulated with ZHAireS-RASPASS and passing at an altitude $h=36\unit{km}$ a.s.l. with $\theta = 90^\circ, 92.5^\circ$ and $95^\circ$. The showers are intercepted at the slant depth of the detector corresponding to showers of different ages: $193\unit{g/cm^2}$ (shower age $s<1$), $880\unit{g/cm^2}$ (shower age $s\simeq1$) and $16550\unit{g/cm^2}$ (shower age $s\gg1$) respectively. The direction $X$ is parallel to ground. By definition, the shower axis passes through $X = 0\unit{km}$ in all cases. The distributions are shown in normal simulation conditions (solid lines), and removing artificially all muons with an energy below $1\unit{TeV}$ (dashed histograms).} \sergio{Only particles reaching the position of the detector within $5\unit{\mu s}$ of a reference particle traveling at the speed of light are considered. The total number of electrons and positrons reaching the detector plane in normal simulation conditions is $\sim 9\times10^{4}$, $\sim 2.2\times10^{5}$ and $954$ for $\theta =90^\circ$, $92.5^\circ$ and $95^\circ$ respectively}.}
    \label{fig:Lateral_Projection2}
\end{figure}

\section{Summary and conclusions}
\label{sec:conclusions}

In this work we have presented the first full three-dimensional dynamic simulation of atmospheric-skimming (AS) showers that develop in the atmosphere with the shower axis not intercepting ground level (figure\,\ref{fig:RASPASSGeometry}). This new class of above-horizon events have been detected with radio pulses at the balloon-borne ANITA detector flying over Antarctica \cite{ANITA:2016vrp, ANITA:2018sgj, ANITA:2020gmv}. We have performed a detailed study of the characteristics of atmosphere-skimming showers showcasing that comprehensive and reliable Monte Carlo simulations of these events are needed to interpret these measurements accurately, and are now feasible using ZHAireS-RASPASS. 

The peculiar geometry of AS showers, in combination with the propagation along a rarefied atmosphere, gives rise to important differences with respect to regular downward-going showers. Depending on their zenith angle AS showers can cross the whole atmosphere without reaching shower maximum (figure\,\ref{fig:PhaseSpace}). Only for sufficiently large $\theta$ the showers can be fully developed before reaching a balloon-borne detector at the nominal altitude of ANITA at $h=36\unit{km}$ (left panel of figure\,\ref{fig:PhaseSpace}). In contrast, a detector at lower altitude is more likely to observe horizontal AS events and even downward-going showers with $\theta<90^\circ$ (right panel of figure\,\ref{fig:PhaseSpace}) not intercepting ground. The altitude of the detector significantly influences its sensitivity to AS shower events, and detailed simulations would be required to asses the exposure of a balloon-borne detector whose altitude changes with time.

Due to the low density in the layers of the atmosphere where AS showers propagate they can travel very long distances, on the order of hundreds of km (figure\,\ref{fig:LongDev}), to be compared with the much smaller distances traveled by downward-going showers on the order of a few km. Notably, the elongation rate and fluctuations in $X_\mathrm{max}$ seem to be in line with those of regular showers when expressed in $\mathrm{g\,cm^{-2}}$ (figure\,\ref{fig:LongDev}). However, fluctuations from one event to another of only a few $\mathrm{g\,cm^{-2}}$ result in differences of tens of kilometers in the location of maximum development.

The propagation in the rarefied layers of the atmosphere also alters the balance between the energy channelled into the electromagnetic and muonic components, and as a consequence, these showers typically have more \textit{invisible} energy than downward-going ones, with the amount of energy not going into the electromagnetic component increasing when $\theta$ is closer to $90^\circ$ and the shower develops in a less dense atmosphere (figure\,\ref{fig:Elongation_rate}). 

The deflection of particles in the magnetic field and propagation in a rarefied medium, induce multiple counting of low energy particles in the longitudinal profile. These particles can in fact cross several times a plane at a fixed depth as they are deviated by the magnetic field in the low-density layers of the atmosphere propagating with little energy loss and attenuation (figure\,\ref{fig:LongDev_Bfield}). This effect is particularly severe for the horizontal (parallel to ground) configuration of the magnetic field for which charged particles typically move away from ground entering the upper and smallest density layers of the atmosphere where they barely lose energy or interact. This calls for a novel representation of the longitudinal profile of these showers that avoids this effect and that we have adopted in this work, consisting on counting particles in bins of the time at which the particles are present in the shower. 

The combined influence of magnetic field deflections in the low-density layers of the atmosphere and the propagation over extensive distances along the shower axis, tends to spread out air shower profiles very significantly in the direction transverse to the shower axis perpendicular to the Earth's magnetic field, as illustrated in figures \ref{fig:Lateral_muons} and \ref{fig:Lateral_electrons}. This has the effect of  \textit{flattening} AS showers that exhibit a larger transverse dimensions along the direction orthogonal to the magnetic field compared to the perpendicular to it. The dynamic nature of the shower and the long distances along which electrons and positrons can gyrate not only alters the distribution of particles in space, but also in time, as shown in figures \ref{fig:Lateral_electrons_time} and \ref{fig:Lateral_Projection}. 

\sergio{The long distances over which AS air showers develop could also open the door to measuring the particle content in the cascade at various stages of its development \cite{Krizmanic:2023hvf}. 
As shown in figure \ref{fig:Lateral_Projection2}, for shower geometries such that the cascade is intercepted in the first stages of its evolution (shower age $s\lesssim 1$), the electromagnetic component is mainly sourced from the decay of $\pi^0$'s produced in early hadronic interactions. On the other side, muon decay is expected to contribute significantly to the number of $e^\pm$ at the detector in those shower geometries where the cascade is intercepted at ages $s\gg 1$.} 

 The peculiar characteristics of the longitudinal and lateral development of AS showers presented in this work \jaime{have been shown to influence the properties of optical Cherenkov emission in this type of showers \cite{Cummings:2021bhg},} 
 and they are expected to have an important influence on the properties of the radio pulses in the MHz - GHz frequency range as well \cite{Zas:1991jv, Alvarez-Muniz:2022uey}. Detailed modelling of these pulses \cite{Tueros_Radio_ICRC2023} is crucial for the interpretation of the signals recorded at balloon-borne experiments exploiting the radio technique. This will be addressed in detail in a followup work.

\section{Acknowledgments} 
This work has received financial support from
Xunta de Galicia, Spain (CIGUS Network of Research Centers,
Consolidaci\'on 2021 GRC GI-2033 ED431C-2021/22 and 2022 ED431F-2022/15),
Feder Funds,
Ministerio de Ciencia, Innovaci\'on y Universidades/Agencia Estatal de Investigaci\'on, Spain
(PID2019-105544GB-I00, PID2022-140510NB-I00, PCI2023-145952-2),
and European Union ERDF.

\bibliography{main_jcap}
 
\end{document}